# How different age groups responded to the COVID-19 pandemic in terms of mobility behaviors: a case study of the United States

**Aliakbar Kabiri [1], Aref Darzi [2], Weiyi Zhou [3], Qianqian Sun [4], Lei Zhang [5, *]**

Department of Civil and Environmental Engineering [1]
University of Maryland
1173 Glenn Martin Hall, College Park, MD 20742
United States
kabiri@umd.edu

Department of Civil and Environmental Engineering [2]
University of Maryland
1173 Glenn Martin Hall, College Park, MD 20742
United States
adarzi@umd.edu

Department of Civil and Environmental Engineering [3]
University of Maryland
1173 Glenn Martin Hall, College Park, MD 20742
United States
qsun12@umd.edu

Department of Civil and Environmental Engineering [4]
University of Maryland
1173 Glenn Martin Hall, College Park, MD 20742
United States
wyzhou93@umd.edu

**Corresponding Author [5, *]**
Lei Zhang, Ph.D.
Herbert Rabin Distinguished Professor
Director, Maryland Transportation Institute
Department of Civil and Environmental Engineering
University of Maryland
1173 Glenn Martin Hall, College Park, MD 20742
United States
Email: lei@umd.edu
Phone: +1 (301) 405-2881

## Abstract

The rapid spread of COVID-19 has affected thousands of people from different socio-demographic groups all over the country. A decisive step in preventing or slowing the outbreak is the use of mobility interventions, such as government stay-at-home orders. However, different socio-demographic groups might have different responses to these orders and regulations. In this paper, we attempt to fill the current gap in the literature by examining how different communities with different age groups performed social distancing by following orders such as the national emergency declaration on March 13, as well as how fast they started changing their behavior after the regulations were imposed. For this purpose, we calculated the behavior changes of people in different mobility metrics, such as percentage of people staying home during the study period (March, April, and May 2020), in different age groups in comparison to the days before the pandemic (January and February 2020), by utilizing anonymized and privacy-protected mobile device data. Our study indicates that senior communities outperformed younger communities in terms of their behavior change. Senior communities not only had a faster response to the outbreak in comparison to young communities, they also had better performance consistency during the pandemic.

## Keywords

COVID-19; Mobile Device Location Data; Mobility Pattern; Social Distancing

# 1. Introduction

In December 2019, a novel coronavirus (COVID-19) began spreading across the world. As of May 18, there have been more than 1,560,000 confirmed cases and more than 94,000 have lost their lives to disease. Non-Pharmaceutical Interventions (NPI) have been considered one of the most effective strategies in fighting against this outbreak in the absence of vaccinations. Therefore, on March 13, the U.S. government declared a national emergency followed by the Coronavirus Guidelines for Americans on March 16, which highlighted the importance of the social distancing practices. In addition to the national emergency, 42 states across the U.S. instituted stay-at-home orders by mid-April 2020 to reinforce the preventive measures. Despite all the efforts conducted to flatten the curve, COVID-19 continues to spread at an alarming pace, which indicates the importance of appropriate NPI planning, especially during the reopening phase.

Earlier studies revealed the linkage between the transmission of COVID-19 and socio-demographic characteristics at both the community and individual level. With a specific focus on the work status variable by difference age groups, a study carried out by Papageorge et al. found that younger workers were more likely to begin teleworking while older workers tend to not work instead (Papageorge, et al., 2020). In another study, which leveraged novel GPS location data, Engle's team analyzed the influence of COVID-19 and social distancing restrictions on individual mobility behavior. The results suggested that counties with larger shares of population over 65 are more responsive to pandemic prevalence and restriction orders (Engle et al., 2020). Some research discussed the differences in social distancing behavior between different socio-demographic groups (Fang, et al., 2020; Lee, et al., 2020; Sun, et al., 2020). A preliminary analysis indicated that states in the U.S. differ in mobility metrics, such as percentage of staying home, miles traveled per person, and percentage of out-of-county trips, when grouped by income levels or population density (Lee, et al., 2020). A more detailed analysis by zooming in on different communities in the United States was conducted by the authors, which showed notable disparities between the mobility patterns of people from different income levels (Sun, et al., 2020). Many studies have shown that being elderly is one of the high-risk phenotypes of COVID-19 (Raifman M and Raifman J, 2020; Covid, C. D. C., & Team, R., 2020; Liu, 2020, Yancy, 2020; Lewnard and Lo, 2020) and age structure is closely related to the progression of the pandemic (Dowd, et al., 2020). Therefore, we aim to investigate the mobility behavior by age group during the pandemic. Related research shows that elderly people have fewer close contacts with non-household members based on an online survey (Canning, et al, 2020). A second study analyzed the connection between age, social distancing behavior, and the progression of the pandemic (Zhang, et al., 2020). Another study treated age as an important consideration when analyzing social contact behavior (Block, et al., 2020).

This research explored the disparities in response to mobility-related, non-pharmaceutical interventions between communities of different age groups. Several aspects of mobility patterns including percentage of people staying at home and social distancing index, have been investigated in this paper to illustrate how different age groups behave during the pandemic. Emergence of mobile device location data has helped many researchers to study the mobility behavior of individuals across the world. In this study, we utilized a panel mobile device location data integrated from several data vendors. This integrated dataset consists of more than 100 million devices across the U.S. and enabled us to calculate several mobility metrics at census tract level using previously developed algorithms (Zhang, et al, 2020). We performed statistical methods, including pairwise t-test and post-hoc analysis, to investigate the differences in temporal mobility behavior patterns between different age groups, which are derived from the mobile device location data. In the following sections, we first describe the datasets used for our study and explain the methodology used for our analysis (Section 2); then, we illustrate findings from our analysis (Section 3); and finally, we discuss conclusions and future directions (Section 4).

## 2. Material and Method
### 2.1. Data Description
#### 2.1.1. Socio-demographic data

We collected age and population data from the American Community Survey (ACS) 5-year estimate (2014-2018) at census tract level across the United States. This data was indirectly downloaded from the NHGIS website, cited in the reference section.

#### 2.1.2. Mobility metrics

Mobility metrics are the key feature in this study. The research team created a robust and representative data panel by integrating multiple mobile device location data sources, which capture the person and vehicle movements. After creating the data panel, we went through comprehensive data cleaning steps considering consistency, accuracy, completeness, and timeliness of the observations in the data panel. We then identified activity locations using the spatio-temporal clustering of raw location data. This step of identifying activity location mainly helps infer the residential and work location of anonymized devices at the census block group level with privacy protection taken into account. After home and work identification, we extracted trips from raw location points using a previously developed recursive algorithm (Zhang, et al, 2020). The trip identification algorithm provides trip-related information, including trip origin and destination, departure time and arrival time, and travel distance of each trip. To estimate the percentage of people staying at home, we considered all anonymized devices with no trips made longer than one mile from their designated residential location as devices staying at home for that calendar day. The last step was to employ a multi-level weighting algorithm to expand the sample observed in our mobile device data to the population level. The final results of our computational algorithms were extensively validated using several independent data sources, including the National Household Travel Survey (NHTS) and American Community Survey. Additionally, the algorithms were peer-reviewed by an external expert panel (Zhang, et al, 2020).

Basic mobility metrics, including percentage of people staying at home, number of trips per person per day, number of work trips and non-work trips per person per day, miles traveled per person per day, and percentage of out-of-county trips per day, were calculated using the aforementioned method. It is worth mentioning that the percentage of people staying home is defined as the number of people without any trip longer than one mile in a day. However, as none of these metrics could represent the different aspects of human mobility patterns, a single metric is introduced to portray the extent of social distancing practices for each census region. The social distancing index (SDI) was calculated as a score-based index to measure the accordance of both residents and visitors to social distancing guidelines (Pan, et al., 2020). A score between 0 and 100 has been assigned to each region by considering the temporal changes in five basic mobility metrics, e.g., percent of staying at home, daily work trips, daily non-work trips, trip distance, and percent of out-of-county compared to baseline days before the start of the COVID-19 pandemic. The research team designed a weighting scheme for integrating the five metrics by considering the importance of each metric based on both conceptual guidelines of social distancing definition and real-world observations.

Table 1 shows the descriptive statistics of the metrics used in this paper.

| Metric | Min | Median | Mean | Max |
| --- | --- | --- | --- | --- |
| % Staying home | 0 | 24.32 | 26.67 | 100 |
| Trips/person | 0 | 3.11 | 3.11 | 6.52 |
| Work trips/person | 0 | 0.45 | 0.48 | 1.81 |
| Non-work trips/ person | 0 | 2.63 | 2.63 | 5.45 |
| Miles travelled/person | 0 | 27.15 | 30.30 | 102.79 |
| % Out-of-county trips | 0 | 16.25 | 19.26 | 100 |
| Social distancing index | 0 | 32.5 | 34.71 | 100 |

**Table 1. Descriptive statistics of the mobility metrics.**

### 2.2. Methodology

In this study, the community categories are defined based on the median age of census tracts. We first sorted the census tracts by median age in ascending order and then picked the top and bottom 20% as young and senior communities, respectively. The performance of following social distancing for both groups are quantified as the percentage change of mobility metrics between the study period (weekdays from March 1 to May 15, 2020) and predefined baseline (weekdays in January and February 2020). The base value of each mobility metric of each census tract will be the mean value of the metric in the baseline period. Utilizing the daily percentage changes for both groups, a t-test was conducted to evaluate the significant difference between two communities, and consequently reveals the mobility difference between different age groups toward the outbreak of the pandemic. Communities have different inherent characteristics that lead to distinct mobility patterns, thus comparing the absolute value of mobility metrics does not fully illustrate their response to the pandemic. Therefore, as we were most interested in capturing the changes in behavior among different age groups during the pandemic, we examined the temporal changes of the mobility metrics compared to the pre pandemic baseline. As an example, young people may have a higher rate of work trips in comparison to the elderly, and this is their internal characteristic. On the other hand, if we only consider the percentage changes of each parameter, it shows the exact effort that each group exerts to prevent the pandemic. In other words, percentage changes can help us capture the exact performance that each group has, regardless of the characteristics of each group. These values are completely comparable.

In addition to the aforementioned day-to-day comparison, similar to the research conducted by Maryland Transportation Institute (MTI), we contrasted the differences in five significant time intervals named 'Pre-Pandemic' (March 1 to March 13), 'Behavior Change' (March 14 to March 23), 'Government Orders and Holding Steady' (March 24 to April 13), 'Quarantine Fatigue' (April 14 to April 24), and 'Partial Reopening' (April 25 to May 15); we then evaluated the potential differences in social distancing performance of two senior and young communities in these time intervals. Moreover, understanding the reaction time to mobility interventions such as the national emergency declaration in different groups, as well as how well they maintain social distancing during the pandemic, are some important questions that need to be addressed. We utilized a post-hoc, pairwise t-test to capture the rate of changes within the groups on different days.

3. Results
   3.1. Temporal change in percentage changes of mobility metrics

In this section, we analyzed the temporal changes in percentage changes of two mobility metrics, namely, percentage of people staying home and the social distancing index from March 1 to May 18, 2020. We calculated the percentage changes in each weekday of the study period in comparison to a baseline before the pandemic, when everything was as usual without any external treatment related to the pandemic, such as the national emergency declaration or stay-at-home order. The baseline is defined as all the weekdays in January and February 2020. It is worth mentioning that only the values of weekdays are considered. The value of mobility metrics of each age group is the daily mean value of all the census tracts in that group, while for the baseline, it is the mean of all weekdays in January and February.

Based on the daily percentage changes of each mobility metric, two graphs (Figure 1) are produced to show the temporal changes of each age group.

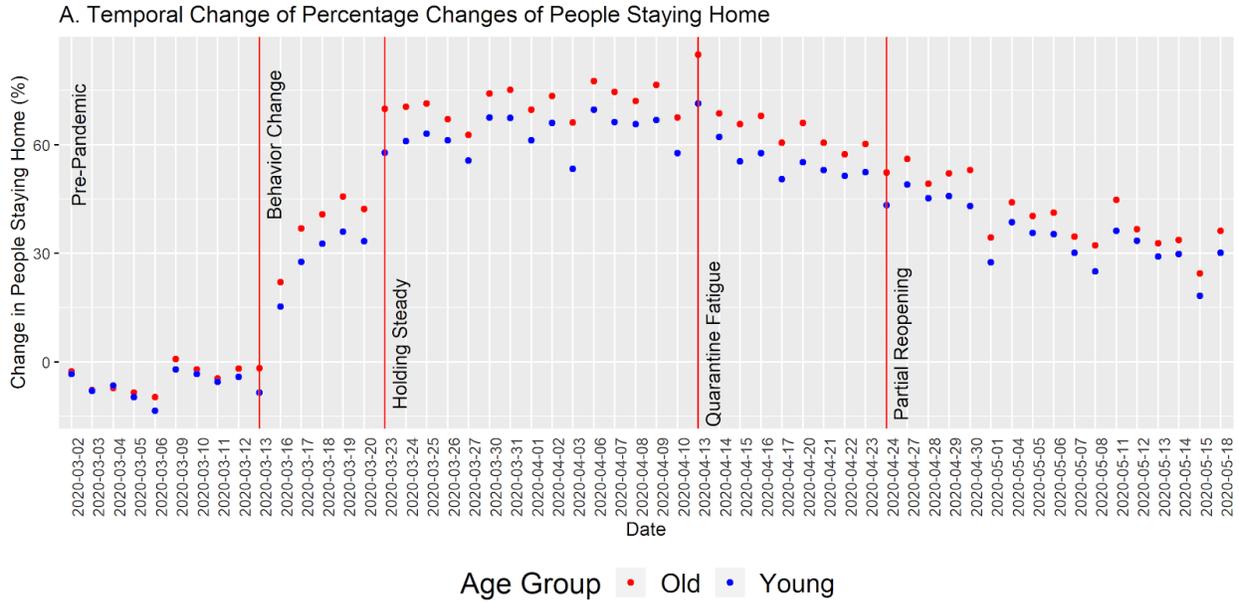

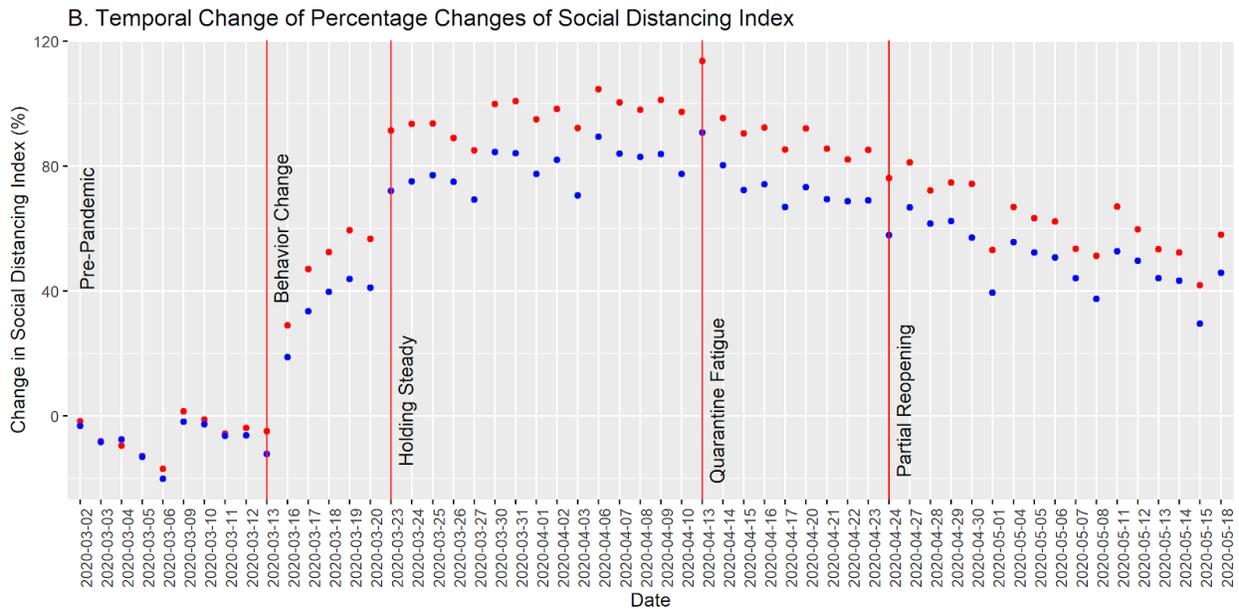

**Figure 1. Temporal trend of daily percentage change for two mobility metrics.** (A) Social Distancing Index (B) and Percentage of People Staying Home in comparison to the baseline (January and February 2020). The red lines represent the boundary between each study period, which is defined in the figure and is discussed in Section 3 and 4.2.

Based on Figure 1, after the national emergency declaration by the federal government on March 13, both senior and young communities began performing social distancing in both aspects of mobility. They continued improving their performance until mid-April 2020 and then stopped enhancing their behavior. This phenomenon is called social distancing inertia (Ghader, et al., 2020). After this time, the trend shifts backwards, indicating that people—no matter what group they belong to—stop improving and even violate social distancing.

By digging more into each age group, we saw that in terms of percentage changes of people staying home, senior communities have higher values in all days (Figure 1.a). In other words, the senior communities outperform the younger communities in staying at home. This finding is crucial, since the literature indicates that older people are more likely to get sick by COVID-19 in comparison to younger people. Considering these two points, we will have a good gauge for moderating the harsh sequences of the outbreak of COVID-19. The same pattern was observed for the social distancing index (Figure 1.b), which also supports that senior communities contribute more in performing social distancing.

### 3.2. Differences between percentage changes in different study periods

In addition to the temporal change of mobility metrics, a t-test method is adopted in this section to statistically analyze the significance of difference between two groups. Based on a day-to-day comparison of the mean values of both metrics, we have found that in almost half of the days before the national emergency declaration the difference is not that significant, as shown in Figure 1. On the other hand, in the days after March 13, the difference between young and old groups is statistically significant. This test confirms that there is a difference between the behavior of senior communities and young communities against the pandemic. Older people are more likely to take better actions in this situation.

In addition, in each time period discussed in Section 3, a t-test is conducted to show whether there is any statistically significant difference between the behavior changes of the young and senior groups, of which the null hypothesis of the t-test is defined as no significant difference (Table 2). As indicated by the results, in all time periods, there are significant differences between both young and senior groups.

| Metric | Study Period | Mean of Young Group | Mean of Senior Group | t-value | Significant? (P-value <0.05) |
| --- | --- | --- | --- | --- | --- |
| Percentage of People Staying Home | Pre-Pandemic | -6.048 | -4.597 | -12.73 | YES |
| | Behavior Change | 36.465 | 44.164 | -33.21 | YES |
| | Government Orders and Holding Steady | 68.047 | 74.198 | -35.63 | YES |
| | Quarantine Fatigue | 57.272 | 63.681 | -27.44 | YES |
| | Partial Reopening | 37.744 | 41.777 | -23.10 | YES |
| Social Distancing Index | Pre-Pandemic | -8.355 | -7.033 | -11.06 | YES |
| | Behavior Change | 43.856 | 57.176 | -55.44 | YES |
| | Government Orders and Holding Steady | 85.193 | 100.701 | -91.62 | YES |
| | Quarantine Fatigue | 74.292 | 89.686 | -70.58 | YES |
| | Partial Reopening | 52.646 | 63.145 | -64.83 | YES |

**Table 2. T-test results of each study period.** This table shows how people in different groups behave over time and demonstrates the effort that each age group puts on performing social distancing in comparison to each other.

Furthermore, Figure 2 shows the distribution of social distancing index of the two age groups in each time period and how it changes over time. It is obvious that in the Pre-Pandemic period, the two groups are similar to each other, yet later, a significant gap appears between the groups. Moreover, they differ in spread. Older people have a more dispersed distribution towards higher values, meaning that overall, they outperform the younger group.

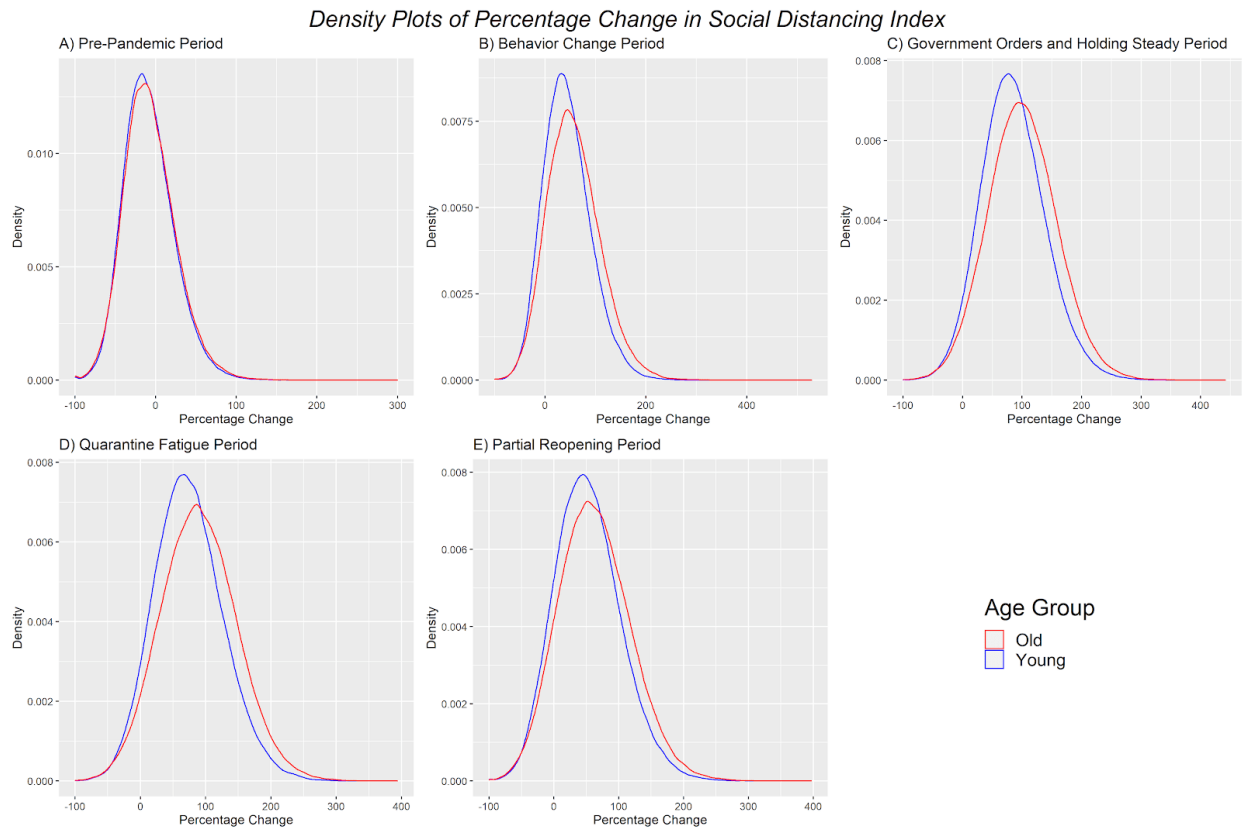

**Figure 2. Distribution of percentage changes in Social Distancing Index (SDI) of the elderly and young tracts in different time periods.**

### 3.3. Post-hoc analysis within each age group

In this section, we investigated the within group changes in young and senior communities over time. We employed pairwise two-sample t-tests to compare the metrics' value day by day. To offer a more detailed look into the differences, we chose the top 150 and bottom 150 census tracts in terms of median age and performed a post-hoc analysis on each pair of the days within each group on both metrics discussed in the previous section—the percentage of people staying home and social distancing index—to see where the difference comes from. Our results indicated significant differences in clusters of consecutive days. Figure 2 shows the results of these census tracts.

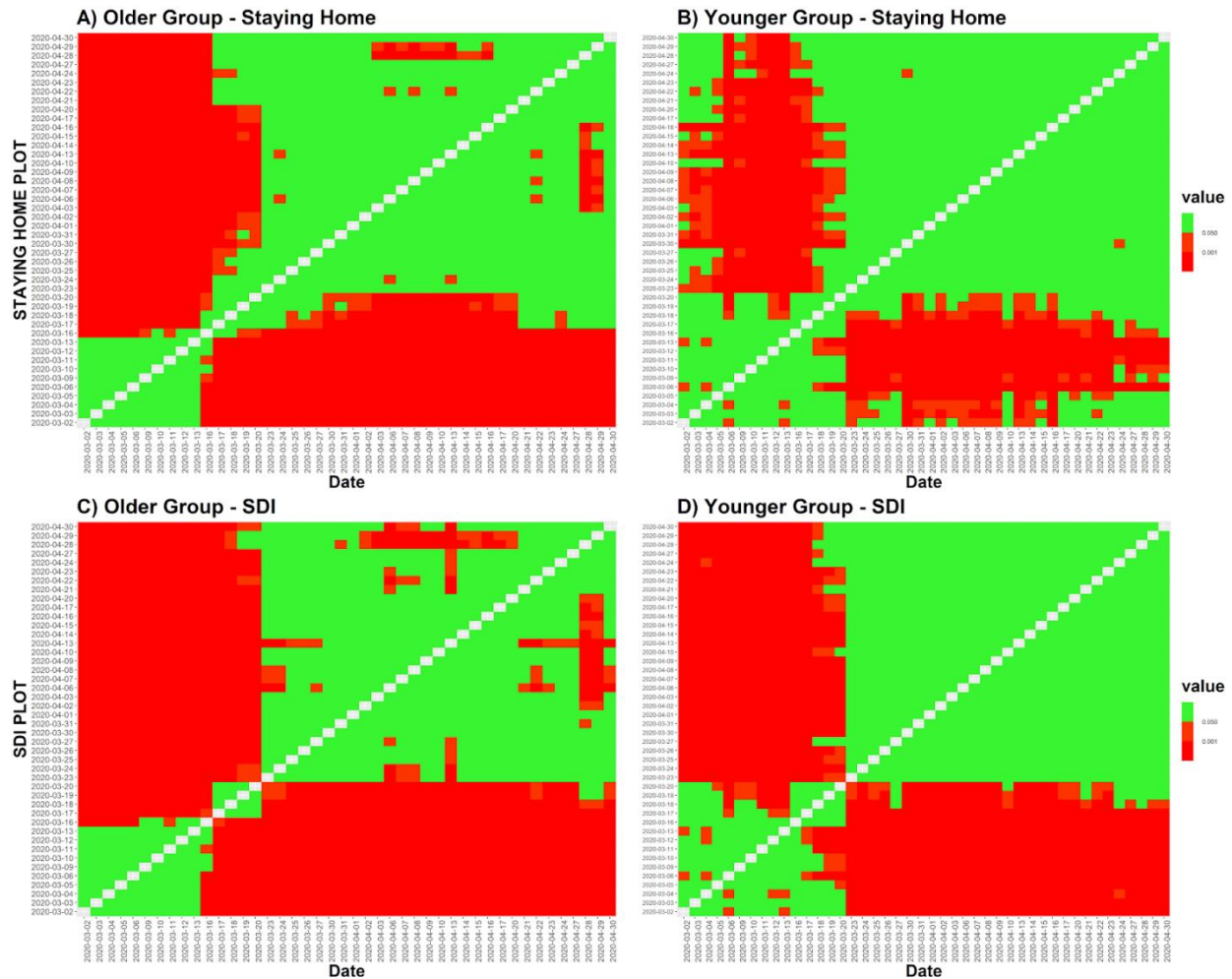

**Figure 3. Pairwise t-test results for two different metrics within each age group**. The p-values of each pair of days are shown. Green cells show no significant difference (P-value greater than 0.05), and red cells show significant differences. (P-value less than 0.05 and even most of them almost near zero.)

Based on Figure 2, the percentage of people staying at home were similar among the two groups from March 1 to March 13. However, a sudden shift was observed in the following days for the older age group. This observation indicates quicker response to the pandemic and national emergency declaration in the elderly group compared to the young group. This indicates that this group is responding to the pandemic and national emergency declaration very fast. On the contrary, in the younger group we saw that even until March 20, the pandemic and the executive order were not significantly considered. Still, until March 20 there are a lot of green cells, which indicates that there is not that much difference between the younger group's behavior before and after the national emergency declaration. Therefore, we can conclude that senior communities were more responsive to the mobility interventions. Furthermore, our analysis revealed that senior groups were more consistent in terms of changing their behavior and practicing social distancing. We observed continuous red cells for the elderly group, which indicates a significant change to their pre-pandemic behavior, while several fluctuations were noticed by the young group. Thus, it concludes that senior communities maintain social distancing better than younger communities. The next two graphs (Figure 2.C and 2.D) also confirm that there is a difference between the response time of the older group and the younger group. It is worth mentioning that the SDI value is made up of many mobility metrics that are discussed in the data description. Therefore, its value is very important and considers mobility behavior of different regions from many different aspects, and this makes it more trustworthy.

## 4. Discussion and Conclusion

The COVID-19 pandemic crisis has affected the daily lives of almost all people. In order to prevent the rapid transmission of the disease, people began changing their mobility patterns by staying home and having less trips voluntarily, or through different restrictions imposed by state and federal governments. Because different people have different characteristics and consequently different behaviors, communities might have distinctive responses to the pandemic. In this paper, we quantitatively investigated the differences between the mobility patterns of senior communities and young communities using privacy-protected, anonymized mobile device location data by comparing the percentage change of mobility metrics, such as social distancing index (SDI) and percentage of people staying home, during the pandemic in comparison to the days before the pandemic to capture the different behaviors of these two groups. The results show that after the national emergency declaration started, both younger and older communities performed social distancing until mid-April, but that as time goes by, they do not maintain their previous social distancing efforts as they did in the first days of the pandemic. Furthermore, our analysis shows that elderly groups reacted to the pandemic faster than young groups. As soon as the national emergency declaration was deployed, these people changed their behavior by staying home and maintaining a higher value of social distancing index. Also, elderly people had a constant performance and kept up their efforts in almost all days. On the other hand, young communities show a disparity in how they behave on different days during the pandemic. Furthermore, we analyzed the overall performance of both groups by using a day-by-day analysis and also through five different critical time intervals such as pre-pandemic, behavior change, etc., and found that in all of these time intervals, there is a significant difference in how these two groups perform social distancing in order to control the spread of the COVID-19.

Future research can shed light on the evolution of the pandemic and how different age groups would return to their normal mobility patterns. The impact of current practices by age group is another future direction that is worth further analysis to understand whether the effort that each group is implementing is enough to prevent the outbreak, or if it is still inadequate. Moreover, this paper can be improved and expanded in many directions. For instance, the current definition of people staying home could be biased towards high-density urban areas. People in urban areas can access shopping in a shorter distance compared to people living in rural areas; thus, it is possible to capture more people staying home in urban areas. Therefore, we can divide each age group into two sub-groups—urban and rural areas—and capture the effect of this factor on age groups.

Finally, this paper reveals that the orders and regulations imposed by policymakers do not result in the same behavior changes in different communities with dissimilar socio-demographic characteristics. For instance, as the results of this paper reveals, stay-at-home orders and travel restrictions have different implications on young and senior communities in terms of mobility behavior shifts. Elderly groups were more agile in adapting to the new guidelines and the changes in their behavior showed a higher compliance rate. Therefore, such data-driven insights from human behaviors are of paramount importance for decision makers to better understand different aspects of their decisions.

**Acknowledgements**

We would like to thank and acknowledge our partners and data sources in this effort: (1) Amazon Web Service and its Senior Solutions Architect, Jianjun Xu, for providing cloud computing and technical support; (2) computational algorithms developed and validated in a previous USDOT Federal Highway Administration's Exploratory Advanced Research Program project; (3) mobile device location data provider partners; and (4) partial financial support from the U.S. Department of Transportation's Bureau of Transportation Statistics and the National Science Foundation's RAPID Program.

**Funding Sources**

This research received partial financial support from the U.S. Department of Transportation's Bureau of Transportation Statistics and the National Science Foundation's RAPID Program.

**CRediT Author Statement**

*Aliakbar Kabiri:* Conceptualization, Methodology, Formal Analysis, Writing – Original Draft, Visualization, Project Administration, Validation, Investigation; *Aref Darzi:* Writing – Original Draft, Data Curation, Investigation, Methodology; *Weiyi Zhou:* Writing – Original Draft, Formal Analysis; *Qianqian Sun:* Writing – Original Draft, Formal Analysis; *Lei Zhang:* Conceptualization, Supervision, Writing – Original Draft, Resources, Funding Acquisition